\documentclass[preprint,5p,twocolumn]{elsarticle}

\usepackage{amssymb}
\usepackage{amsmath}

\usepackage{lineno}
\usepackage{xcolor}

\journal{Astroparticle Physics}

\begin{document}

\begin{frontmatter}



\title{Scintillation characteristics of an undoped CsI crystal at low-temperature for dark matter search}




\author[a,b]{W.~K.~Kim}
\author[c]{H.~Y.~Lee}
\ead{hylee@ibs.re.kr}
\author[b]{K.~W.~Kim}
\ead{kwkim@ibs.re.kr}
\author[d]{Y.~J.~Ko}
\author[b]{J.~A.~Jeon}
\author[e]{H.~J.~Kim}
\author[a,b]{H.~S.~Lee}
\ead{hyunsulee@ibs.re.kr}

\address[a]{University of Science and Technology~(UST), Daejeon, 34113, Republic of Korea}
\address[b]{Center for Underground Physics, Institute for Basic Science~(IBS), Daejeon, 34126, Republic of Korea}
\address[c]{Center for Exotic Nuclear Studies, Institute for Basic Science~(IBS), Daejeon, 34126, Republic of Korea}
\address[d]{Department of Physics, Jeju National University, Jeju, 63243, Republic of Korea
}
\address[e]{Department of Physics, Kyungpook National University, Daegu, 41566, Republic of Korea}

\begin{abstract}
The scintillation characteristics of 1\,g undoped CsI crystal were studied by directly coupling two silicon photomultipliers (SiPMs) over a temperature range from room temperature to  86\,K.
The scintillation decay time and light output were measured using x-ray and gamma-ray peaks from a $^{109}$Cd radioactive source.
An increase in decay time was observed as the temperature decreased from room temperature to 86\,K, ranging from 76\,ns to 605\,ns.
The light output also increased, reaching 26.2 $\pm$ 1.3 photoelectrons per keV electron-equivalent at 86\,K.
Leveraging the significantly enhanced scintillation light output of the undoped CsI crystal at low temperatures, coupling it with SiPMs results in a promising detector for rare event searches. 
Both cesium and iodine have an odd number of protons, making them suitable targets for probing dark matter-proton spin-dependent interactions.
This study evaluates the sensitivity of the proposed detector to such interactions, incorporating the Migdal effect and assuming 200 kg of undoped CsI crystals for dark matter searches.
The results indicate that undoped CsI coupled with SiPM can achieve world-competitive sensitivity for low-mass dark matter detection, particularly in the context of dark matter-proton spin-dependent interactions.
\end{abstract}



\begin{keyword}


Dark matter \sep undoped CsI \sep SiPM \sep Low temperature measurement
\end{keyword}

\end{frontmatter}


\section{Introduction}
\label{sec:intro}

Scintillating crystals are widely employed in the search for rare event interactions such as dark matter~\cite{crystalwimp,Amare:2018sxx,Bernabei:2018yyw,Adhikari:2019off,Zani:2021ije,CRESST:2022dtl} or coherent elastic neutrino-nucleus scattering (CE$\nu$NS)~\cite{Akimov:2017ade,NEON:2022hbk}.
Since event rates are larger at low energy, low threshold detectors are essential for dark matter or CE$\nu$NS detection.
In the case of scintillators, higher light yields are strongly associated with lower energy thresholds.
The introduction of dopants during crystal growth, such as thallium in halide crystals (i.e., CsI(Tl) and NaI(Tl)), enhances radiative recombination and improves light emission efficiency. These doped crystals have been widely utilized at room temperature, coupled with photomultipliers, for dark matter and CE$\nu$NS detection experiments~\cite{Akimov:2017ade,NEON:2022hbk,KIMS:2018hch,COSINE-100:2021xqn,COSINE-100:2021zqh,Amare:2021yyu,sabre2,Bernabei:2020mon}. The highest reported light yields from doped NaI(Tl) crystals were 22 photoelectrons (PE) per keV~\cite{NEON:2022hbk} with a novel crystal encapsulation technique~\cite{Choi:2020qcj}.
There have been reports suggesting that undoped CsI and NaI crystals exhibit higher light yields at low temperatures with PMTs~\cite{amsler2002,park2021,nishimura1995,schotanus1990,1221872,COSINUS:2017bco,Poda:2021hsv} or silicom photomultipliers (SiPMs)~\cite{Ding:2022jjm}.
At the liquid nitrogen temperature of 77\,K, the undoped CsI crystal, coupled with SiPMs and a wavelength shifter, achieved a light yield of 30.1 PE/keV~\cite{Wang:2022ekc}.

We investigate the scintillation characteristics of an undoped CsI crystal to assess their temperature-dependent behavior.
To maximize the light collection efficiency of the undoped CsI crystal, two SiPMs are directly attached to the crystal.
In a temperature range between 86\,K and 293\,K, we measured the light yield, scintillation decay time, and energy resolution.
Due to the exhibition of high dark count rates and crosstalks~\cite{lee2022,Lightfoot:2008im}, we have carefully analyzed and corrected effects of them on the analysis of measurements.

The technique for radiopure CsI crystal growth ~\cite{Kim:2003ms,Kim:2005rr,Lee:2007iq} has been developed for the KIMS experiment, which used CsI(Tl) crystals coupled with photomultipliers (PMTs) for WIMP dark matter search~\cite{Kims:2005dol,KIMS:2007wwj,sckim12}.
The production of 200\,kg undoped CsI crystals can be performed at the Institute for Basic Science (IBS), Korea, which offers a powder purification facility~\cite{Shin:2018ioq,Shin:2020bdq,Shin:2023ldy} and Kyropoulos crystal growers~\cite{Ra:2018kkl}, installed and employed for the growth of radiopure NaI(Tl) crystals~\cite{COSINE:2020egt,Lee:2023jbe}.
Additionally, about 400\,kg of purified CsI powder remnants from the KIMS CsI(Tl) crystal growing have been stored at the Yangyang underground laboratory.
We conducted a feasibility study for rare event searches by measuring the characteristics of a small size undoped CsI crystal. Based on the high light yield observed from these measurements and the low-background CsI crystal growing technology, we evaluate the dark matter detection sensitivities of a future experiment using a detector based on undoped CsI crystals coupled to SiPMs.
In the low-mass dark matter region for the dark matter-proton spin-dependent interaction, this detector can reach a world competitive sensitivity, exploring a new dark matter parameter space.

\section{Scintillation characteristics of the undoped CsI crystal}
\subsection{Experimental setup}
\begin{figure}[htb]
    \begin{center}
        \includegraphics[width=1.0\columnwidth,keepaspectratio]{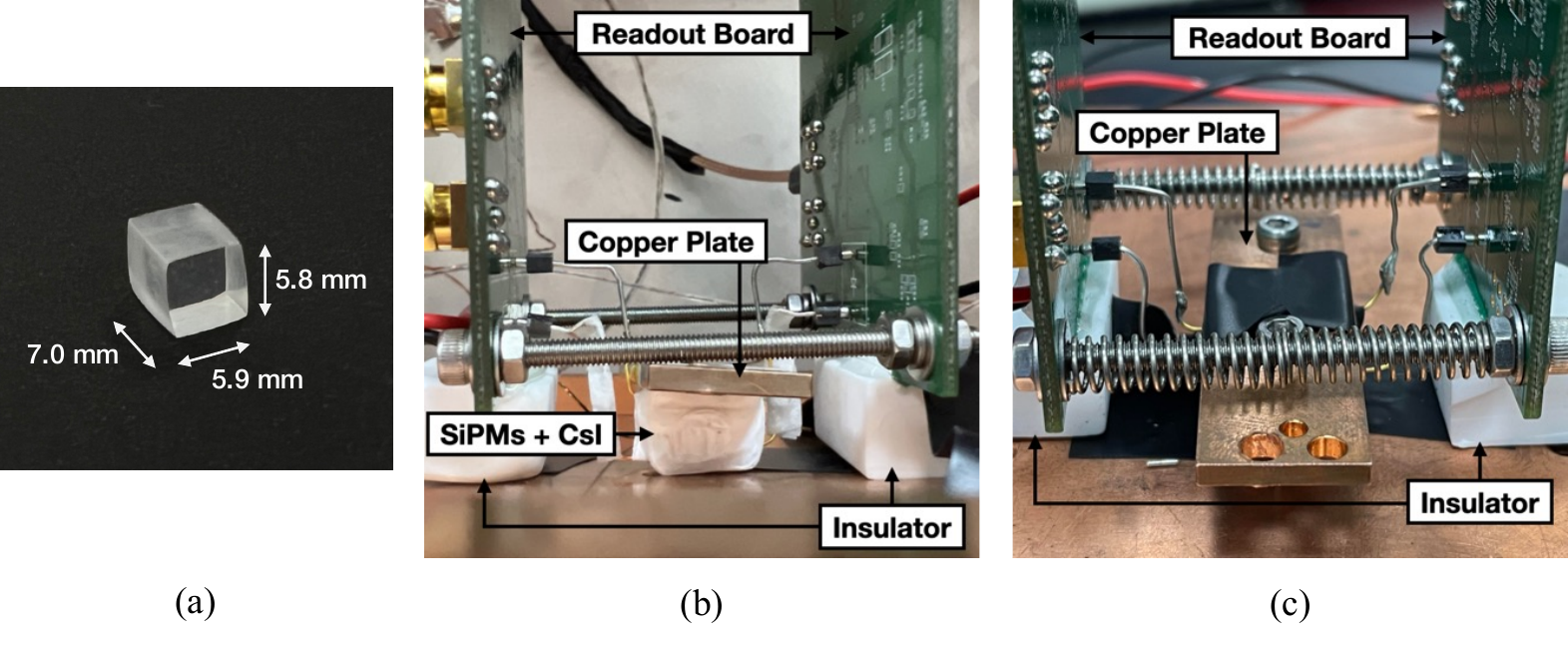}
    \end{center}
		\caption{A 5.8\,mm $\times$ 5.9\,mm $\times$ 7.0\,mm undoped CsI crystal is employed for the experiment (a).  The crystal is coupled with two SiPMs, enclosed with multiple layers of soft polytetrafluoroethylene (PTFE) sheets, and connected to readout boards (b). Crystal and SiPM are in contact with a copper plate for heat transfer, while the readout board is positioned on an insulator to prevent heat transfer to the detector (c). The entire assembly is placed inside a cryostat chamber.}
    \label{setup}
\end{figure}

The detector assembly used in this study consists of an undoped CsI crystal from the Institute for Scintillation Materials (ISMA) in Ukraine and two SiPMs from Hamamatsu Photonics. The undoped CsI crystal has dimensions of 5.8\,mm $\times$ 5.9\,mm $\times$ 7.0\,mm with a mass of 1.0\,g, as illustrated in Fig.~\ref{setup} (a).
Two SiPMs (model number S13360-2172) with an active area of 6.0\,mm $\times$ 6.0\,mm were directly attached to two ends of the crystal.
The crystal and the SiPMs were mechanically attached without any optical coupling due to uncertainties regarding the optical grease and optical pad clarities, which were not guaranteed for temperature below $-$60\,$^{\circ}$C.
This SiPM consists of 14,400 micropixel arrays, each of size 50\,$\mu$m $\times$ 50\,$\mu$m, covered with a quartz window.
The entire assembly was tightly wrapped with soft polytetrafluoroethylene (PTFE) sheets and connected to homemade readout boards (customization available from NOTICE Korea \footnote{http://www.noticekorea.com}) equipped with electronics for signal readout, pre-amplification, and bias voltage supply, as depicted in Fig.~\ref{setup} (b).

\begin{figure}[htb!]
    \begin{center}
    \includegraphics[width=1.0\columnwidth,keepaspectratio]{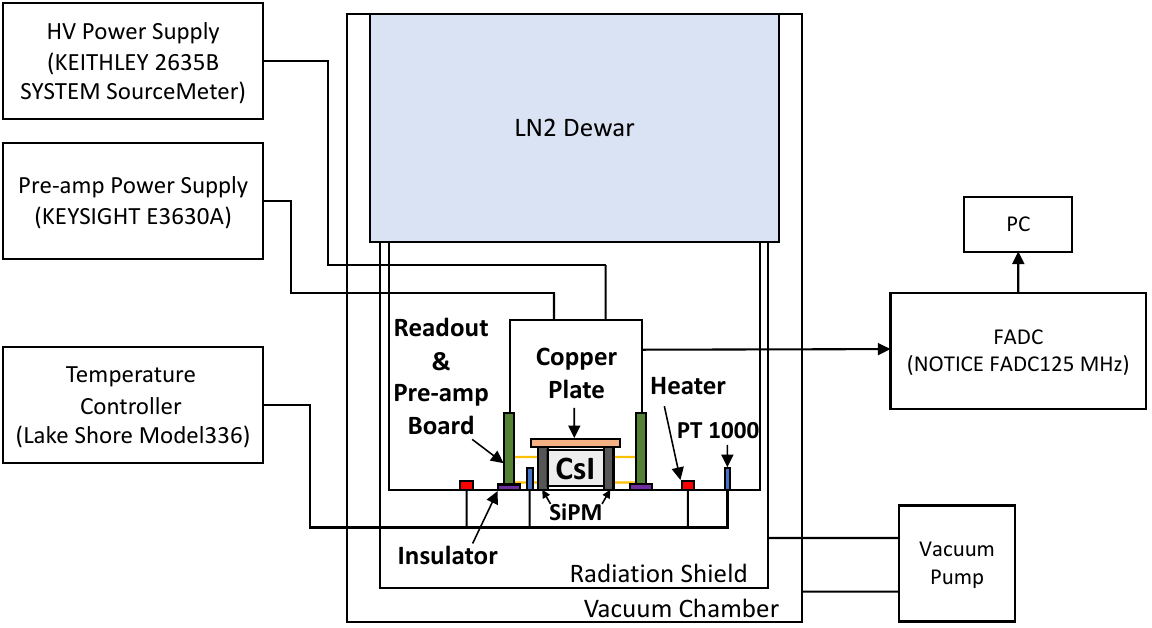}
    \end{center}
    \caption{Schematic drawing of the experimental setup.}
    \label{scheme}
    \end{figure}

The CsI-SiPMs setup was installed inside a cryostat vacuum chamber equipped with a temperature control unit, as depicted in Fig.~\ref{scheme}.
The liquid nitrogen dewar is connected to the copper plate through a thermal link composed of multiple copper wires. The CsI-SiPMs setup was mounted on the copper plate and covered with a radiation shield made of iron to protect against external radiation heat.
Rubber insulators were introduced to prevent the conduction of heat from the readout board, as shown in Fig.~\ref{setup} (b).
To control the temperatures of the CsI crystal, two heaters were incorporated on the copper plate.
Two PT-1000 temperature sensors were employed to monitor the temperature inside the chamber using the temperature control unit of Lake Shore Model 336.
One sensor was placed in close proximity to the crystal, while the other was positioned at the edge of the copper plate.
The temperature control unit regulates the heater to manage crystal (or system) temperature variations, maintaining a stability level of 0.1 degree.
We optimized the thermal connection performance between the nitrogen dewar and copper plate to achieve a minimum temperature closer to 77\,K while ensuring efficient heat transfer to the crystal to increase its temperature when necessary.
In this setup, we successfully reached a temperature of 86\,K with both SiPMs and readout boards in operation.

A KEITHLEY 2635B SYSTEM SourceMeter is employed to provide the SiPM bias voltage through the readout boards, which are operated with a low-voltage supply using a KEYSIGHT E3630A.
The homemade readout electronics amplifies the SiPM signals, and the amplified signals are digitized using a 125\,MHz, 12-bit flash-analog-to-digital converter (FADC) supplied by NOTICE Korea.

\subsection{Measurements}

The performance of the SiPM is affected by the overvoltage, defined as $\Delta V = V_{bias}-V_{breakdown}$, where $V_{breakdown}$ is the breakdown voltage.
Data were collected using six different overvoltages ranging from 3\,V to 8\,V, with a 1\,V interval at room temperature.
The gain, represented by the maximum height of single photoelectron pulses, was evaluated at each overvoltage.
Since the photo detection efficiency (PDE) of SiPM varies with gain or overvoltage~\cite{ferri2014,Teranishi:2021wsl,sze1966,dinu2017,NepomukOtte:2016ktf,dinu2015,collazuol2011}, we maintained a consistent overvoltage for the temperature-dependent measurements.

The temperature range for detector testing spanned from 86\,K to 293\,K (room temperature). 
 Although data were taken across this temperature range, we encountered difficulties stabilizing the temperature at several specific points such as at 120, 250, and 260\,K. These points were therefore excluded from the analysis to avoid introducing uncertainties.
Additionally, due to the limitations of the present setup, complete elimination of heat from the readout board was not achievable, resulting in a minimum temperature of 86\,K, which is above the aimed temperature of 77 K.

\begin{figure}[htb]
    \begin{center}
    \includegraphics[width=1.0\columnwidth,keepaspectratio]{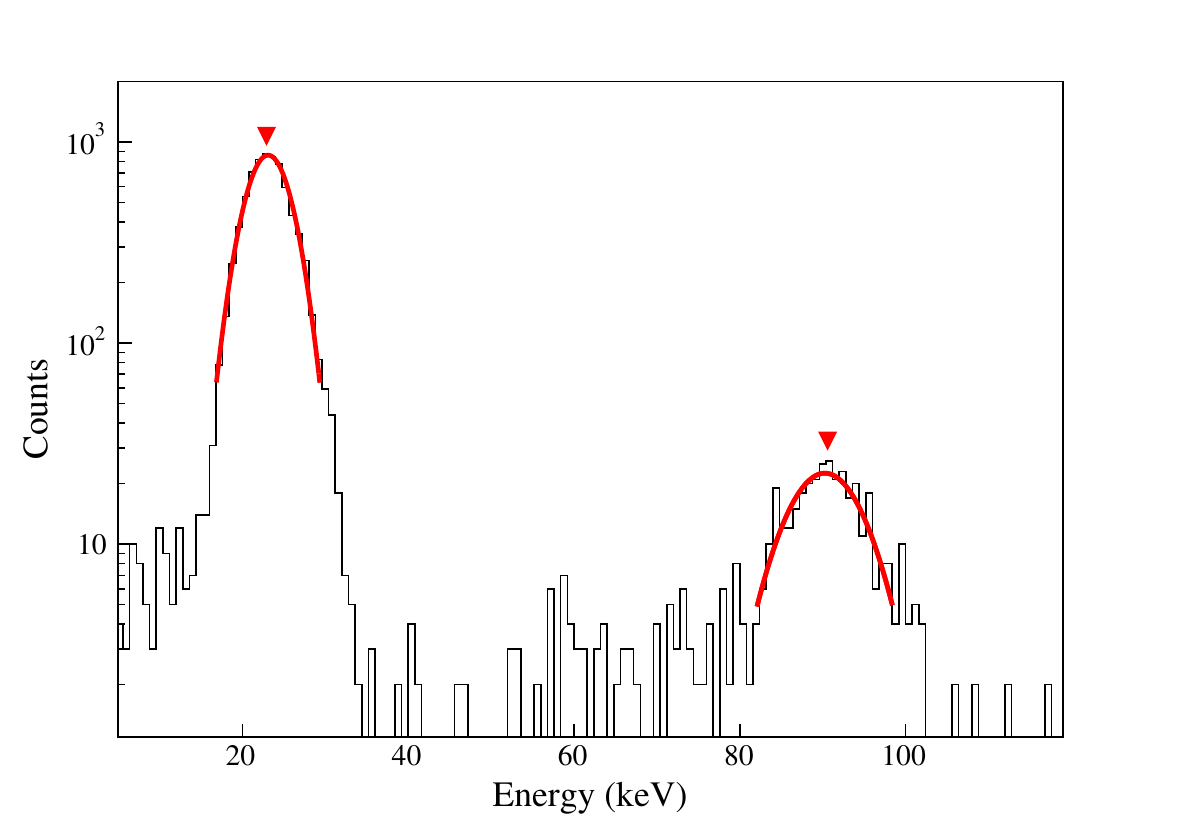}
    \end{center}
    \caption{
    Energy spectrum of the $^{109}\text{Cd}$ calibration source, measured at 150\,K with $\Delta V = 5$ V. The red triangles mark the two energy peaks at 23.0\,keV and 88.0\,keV.
}
    \label{energyspectrum}
\end{figure}

We characterized the scintillation of an undoped CsI crystal using a $^{109}$Cd source that emits 22.1\,keV and 25.0\,keV x-rays as well as an 88.0\,keV $\gamma$-rays.
Because of the energy resolution, the 22.1\,keV and 25.0\,keV x-ray peaks overlapped and were roughly adjusted to around 23.0\,keV, as shown in Fig.~\ref{energyspectrum}.
Energy calibrations were performed using the 88.0\,keV position. At low temperatures, the 23.0\,keV peak is utilized after confirming the linearity of the two peaks. This decision is due to the saturation of the signal corresponding to the energy of the 88.0\,keV peak in our electronics, caused by increasing gain at low temperatures.

\subsubsection{Dark count rate \& Crosstalk}
\label{sec:DCR}

\begin{figure}[htb]
    \begin{center}
    \includegraphics[width=1.0\columnwidth,keepaspectratio]{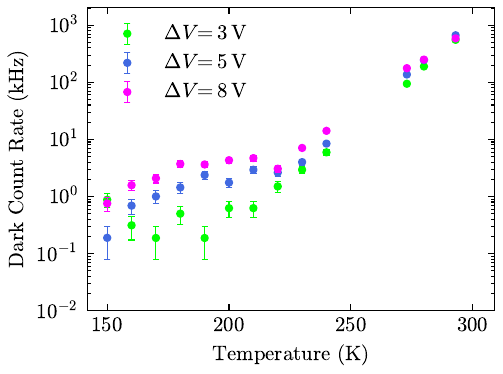}
    \end{center}
    \caption{Dark count rate as a function of temperature for three different overvoltages.}
    \label{dcr}
\end{figure}

Two critical factors in understanding SiPM data are the dark count rate (DCR) and crosstalk.
We measured the DCR and crosstalk using a 1.6\,$\mu$s time window located prior to the event-triggered position in the recorded signal data. This time window, referred to as the pedestal region, is chosen to exclude the influence of the scintillation event and SiPM after-pulses.

The DCR is a significant noise source, particularly at room temperature, while it decreases significantly at lower temperatures~\cite{Lightfoot:2008im,lee2022,Vacheret:2011zza}.
The rate is calculated by dividing the number of clusters, which are isolated pulse typically equivalent to single photoelectrons, identified in the pedestal regions by the corresponding time interval from the offline analysis.
Since the DCR depends on both temperature and bias voltage, we assessed the DCR by varying temperature and overvoltage, as shown in Fig.~\ref{dcr}.
The DCR increases with higher overvoltage and exhibits a dramatic decrease at lower temperatures.
For instance, at room temperature, the DCR is about 700\,kHz with $\Delta V$ = 5\,V and decreased to 0.2\,kHz at 150\,K.
Below 150 K, the DCR drops further, resulting in poor statistical reliability, therefore, data in this region were not included in the analysis.


Crosstalk refers to the probability that a secondary photon, emitted during the avalanche generation process in one pixel following a true photon hit, triggers another pixel, resulting in multiple photoelectron events.
Crosstalk can be categorized into internal Crosstalk (iCT) and external Crosstalk (eCT), depending on where the secondary avalanche occurs.
iCT occurs when the secondary avalanche happens in a nearby pixel within the same SiPM, while eCT occurs when the emitted luminescence escapes from the SiPM and fires another SiPM at a distance~\cite{Wang:2022ekc}.
Crosstalk is independent of temperature variation but highly dependent on bias voltage~\cite{lee2022,Lightfoot:2008im}. 
Both iCT and eCT probabilities were measured at room temperature. 

The iCT probability is estimated from the charge distribution of DCR, as shown in Fig. 3 of Ref.~\cite{lee2022}.
We can easily identify single-photoelectron and multi-photoelectron events to evaluate the crosstalk probability as the ratio of the number of charge-weighted multi-photoelectrons to single-photoelectrons.
This measurement was first performed with a single SiPM turned on.
Then, the same measurement was repeated with both SiPMs turned on. In this case, a higher crosstalk probability was observed compared to the single SiPM measurement, indicating the presence of eCT due to the secondary avalanche in the adjacent SiPM.
The ratio between these two probabilities is used to estimate eCT.
Table~\ref{crosstalk} summarizes the measured crosstalk probabilities at different overvoltages, averaged over the two SiPMs.
Due to the increasing gain of the SiPM with higher overvoltages, the crosstalk probability also rises.





\begin{table}[htb]
    \centering
    \caption{Crosstalk probability of SiPM measured at different overvoltages at room temperature. iCT and eCT represent internal and external crosstsalk respectively.}
            \def\arraystretch{1.3}
	
    \begin{tabular}[t]{ccc}
\hline
\hline
     Overvoltage (V) & iCT (\%) & eCT (\%)\\
 \hline
    3 & 11.9 $\pm$ 0.3 &   7.6 $\pm$ 0.4 \\
    4 & 13.9 $\pm$ 0.3 &   8.8 $\pm$ 0.2 \\
    5 & 27.3 $\pm$ 0.3 &   8.1 $\pm$ 0.2 \\
    6 & 54.5 $\pm$ 0.4 &  10.6 $\pm$ 0.2 \\
    7 & 71.2 $\pm$ 2.2 &  14.2 $\pm$ 0.3 \\
    8 & 90.8 $\pm$ 7.8 &  33.6 $\pm$ 0.3 \\
\hline
    \end{tabular}
    \label{crosstalk}
\end{table}

\subsubsection{Light yield \& Energy resolution}

The light yield of the undoped CsI crystal was measured using the $^{109}$Cd source for temperatures ranging from 86\,K to 293\,K.
To maintain a consistent gain across different temperatures, overvoltage was kept constant.
The results, shown in Fig. 5, are presented alongside previous measurements using typical PMTs~\cite{amsler2002,park2021}, with all three datasets normalized to have the same value at 280\,K.
We show data at 5\,V overvoltage for the temperature-dependent analysis, as measurements at higher overvoltages become less reliable due to increased crosstalk and DCR at elevated temperatures.
The observed increase in light yield at lower temperatures aligns with earlier measurements.
At 86\,K, the total light yields for overvoltages of 3\,V and 5\,V are $21.3 \pm 0.8$ and $22.9 \pm 0.8$ PE/keV, respectively. The highest yield of $26.2 \pm 1.3$ PE/keV was obtained at the same temperature with overvoltages of 8\,V, corresponding to the voltage with the highest gain.
The contributions from DCR and crosstalk are considered in the light yield estimation and the errors associated with these values are included as systematic uncertainties.
The light yield is calculated as the charge integrated in the analysis time window divided by the charge of a single photoelectron (SPE) and the energy of the peak from the radioactive source. Any charge contributed by the DCR should be subtracted from the total charge only to account charges from scintillation events.  Additionally, the crosstalk probability should be corrected to account effective charge produced by single-photoelectron. This process can be described as following,

\begin{equation}
		Light Yield = \frac{Q_{Tot} - Q_{DCR}}{q_{SPE}\cdot (1+p_{iCT})\cdot(1+p_{eCT})}\cdot\frac{1}{E_{cal}},
	\label{eqn:lightyield}
\end{equation}
\begin{align*}
		\qquad Q_{DCR} = (q_{SPE} \cdot \mathrm{DCR} \cdot T),
\end{align*}
where $Q_{Tot}$ is the total charge within the timing window of this analysis (denoted as $T$), which is 16 $\mu$s, $q_{SPE}$ denotes the charge of SPE, and $p_{iCT}$ ($p_{eCT}$) indicates the internal (external) crosstalk probability of the SiPM as summarized in Table~\ref{crosstalk}. $E_{cal}$ represents the energy from the $^{109}$Cd source, either 23.0 or 88.0\,keV.

The emission wavelength of the undoped CsI crystal is peaked at 310\,nm at room temperature and measured to be temperature dependent and equal to 340\,nm at 77\,K~\cite{Woody:1990hq}.
At this wavelength, SiPMs have an almost twice higher PDE of roughly 30\% than that of typical PMTs, which have a PDE of less than 15\%.
This results in an increased light yield compared to PMT measurements~\cite{Ding:2020uxu}.
Similar results were reported by other groups~\cite{Wang:2022ekc}, achieving a light yield of 30.1 PE/keV at 77\,K with SiPMs coupled to undoped CsI crystals and a wavelength shifter from 340\,nm to 420\,nm.

\begin{figure}[htb]
    \begin{center}
    \includegraphics[width=1.0\columnwidth,keepaspectratio]{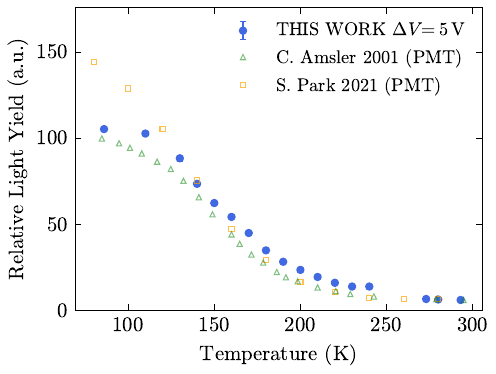}
    \end{center}
    \caption{The relative light yield of CsI, normalized to have same value at 280 K, is depicted as a temperature-dependent function.
    The data are presented for $\Delta V = 5\,V$ (blue filled circles), with a value of $0.88 \pm 0.02$ PE/keV at 280 K.
    Additionally, for comparison, two previous measurements conducted using PMT are overlaid in green open triangle~\cite{amsler2002} and red open square~\cite{park2021}. The overall trend aligns with the results from other measurements.}
    \label{lightyield}
\end{figure}

	

\begin{figure}[htb]
    \begin{center}
    \includegraphics[width=1.0\columnwidth,keepaspectratio]{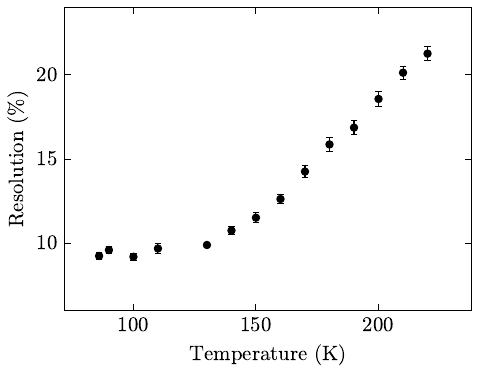}
    \end{center}
		\caption{Energy resolution (Standard deviation/Energy) of 23.0\,keV mixed peak from 22.1\,keV and 25.0\,keV x-rays of the $^{109}$Cd source at different temperatures with $\Delta$V = 5 V.}
    \label{resolution}
\end{figure}

We estimated the standard deviation of the mixed 23.0\,keV peak with a Gaussian fit to evaluate the resolution, calculated as the standard deviation divided by energy.
As shown in Fig~\ref{resolution}, the improved resolution at low temperatures demonstrates the enhancement of light output.

\subsubsection{Decay time}

We investigated the temperature-dependent variation of the scintillation decay time of the undoped CsI crystal with overvoltage of 5\,V for SiPM.
The accumulated waveforms at each temperature were fitted with a single exponential function.\

    \begin{align}
	f(t) = A\exp(-(t-t_{0})/{\tau}),
    \end{align}
where $\tau$ is the decay constant. A and $t_{0}$ represent the normalization factor and the rising edge position of the scintillation, respectively.

\begin{figure}[ht]
    \begin{center}
    \includegraphics[width=1.0\columnwidth,keepaspectratio]{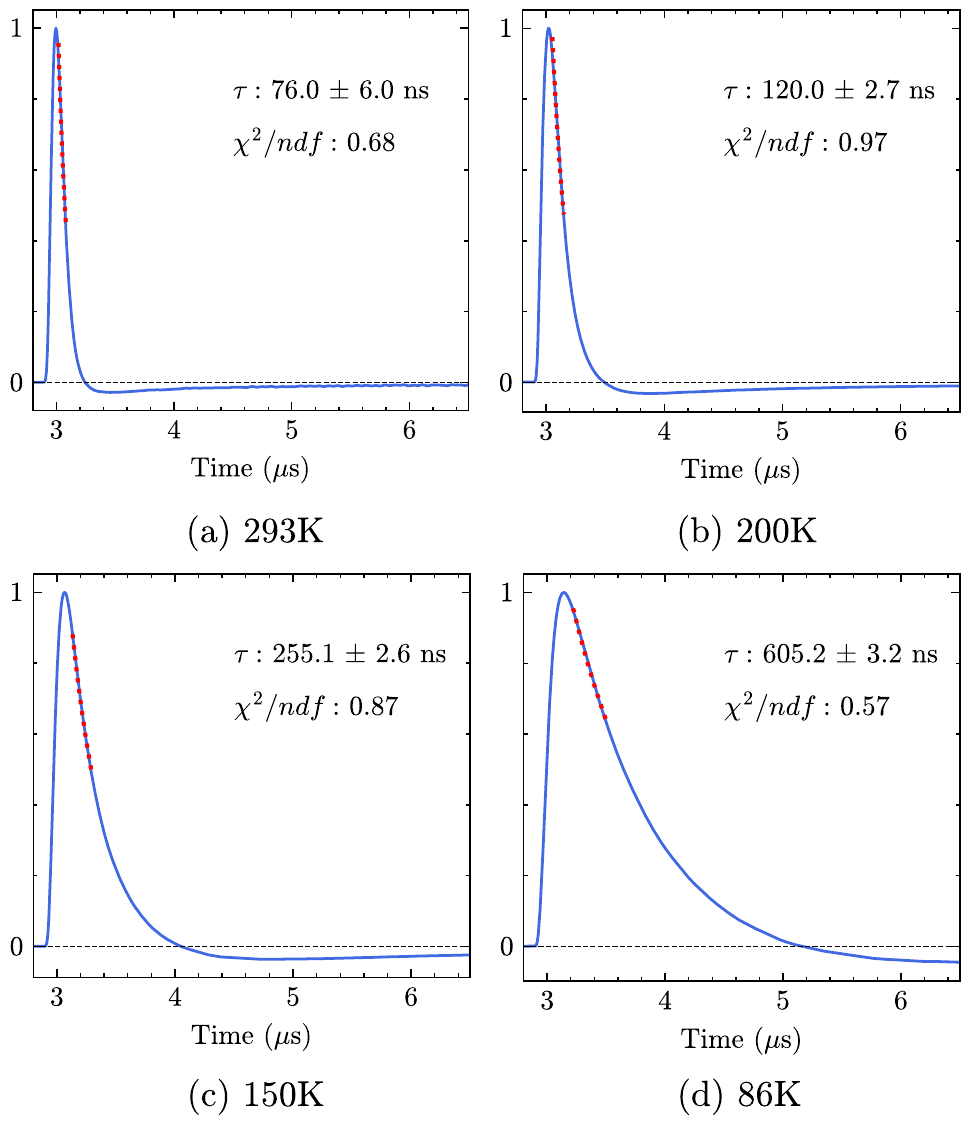}
    \end{center}
    \caption{Accumulated waveforms of the peak from $^{109}$Cd with $\Delta$ V = 5\,V are shown at temperatures of 293\,K, 200\,K, 150\,K, and 86\,K, depicted in panels (a) to (d). The red dashed line represents the fitted exponential function, and the decay constant is described in each plot.}
    \label{decayfit}
\end{figure}

The waveforms and fitted results for the four selected temperatures, presented in Fig.~\ref{decayfit}, revealed an undesired undershoot in the signal generated by our SiPM readout electronics.
The undershoot varied from event to event and appeared to be influenced by both the signal amplitude and baseline fluctuations, likely due to non-idealities in the readout system.
Given the uncertainty in applying a reliable, event-independent correction model, we instead limited the waveform fitting to a short time window and applied a single exponential to avoid misestimating the decay time due to the undershoot, even though previous PMT measurements indicate the presence of two decay components\cite{amsler2002,park2021}.

\begin{figure}[htb]
    \begin{center}
    \includegraphics[width=1.0\columnwidth,keepaspectratio]{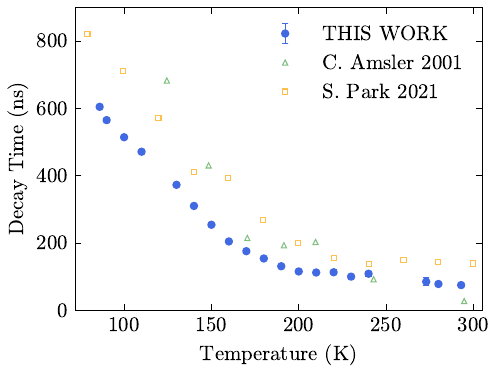}
    \end{center}
    \caption{
    Decay time of a pure CsI crystal as a function of temperature with $\Delta$ V = 5\,V (blue filled circles). Two previous measurements (green open triangle~\cite{amsler2002} and red open squares~\cite{park2021}) using PMTs are overlaid for comparison}
    \label{decaytime}
\end{figure}

The temperature-dependent decay time of the undoped CsI crystal is shown in Fig.~\ref{decaytime} and compared with previous PMT measurements\cite{amsler2002,park2021}.
Our measurements demonstrate a well-agreed trend of increased decay time at low temperatures.
Slightly faster decay times in our measurements at all temperatures result from evaluating the decay time within a short time range, to exclude the severe undershoot previously described.
The measured decay time is 76.0 $\pm$ 6.0\,ns at room temperature and 605.2 $\pm$ 3.2\,ns at 86\,K.
Generally, a longer decay time is advantageous for identifying nuclear recoil events using pulse shape discrimination (PSD).
We plan to test PSD capability with a neutron source at liquid nitrogen temperatures in the future.

\section{200\,kg undoped CsI crystals for the dark matter search}

Following up on the high light yield of the undoped CsI crystal at liquid nitrogen temperature, we consider future dark matter search experiments with 200\,kg of undoped CsI crystals coupled with SiPM arrays, especially to search for low-mass dark matter with spin-dependent interactions.
Both cesium and iodine have non-zero proton spin due to their odd proton numbers, providing an advantage for searching for WIMP-proton spin-dependent interactions.

The low-background CsI crystal growth technique was successfully developed for the KIMS CsI(Tl) dark matter search experiments~\cite{Kim:2003ms,Lee:2007af}.
The primary sources of internal background in CsI crystals were identified as $^{137}$Cs and $^{87}$Rb.
The presence of $^{137}$Cs is mainly attributed to water used in the chemical process of cesium extraction.
By employing purified water, the contamination of $^{137}$Cs can be reduced by one order of magnitude in the CsI powder~\cite{Kim:2005rr}.
The $^{87}$Rb nuclei are extracted using the recrystallization method, taking advantage of the different solubility of RbI and CsI in water~\cite{Lee:2007af}.
About 400\,kg of remnant CsI powder, left over from the KIMS CsI(Tl) crystal growth process, has been stored at the Yangyang underground laboratory for 20 years.
This reserve can be effectively utilized to grow 200\,kg of undoped CsI crystals for future experiments.
Additionally, a powder purification facility based on the recrystallization process was developed for NaI powder~\cite{Shin:2018ioq,Shin:2020bdq,Shin:2023ldy}.
This facility can be re-adapted for CsI powder purification, providing a streamlined process for obtaining low background levels in the CsI crystals if required.

Low-background crystal growth can be conducted at the Institute for Basic Science in Korea, where crystal growers have been developed for low-background NaI(Tl) crystals.
The Kyropulous growers designed for NaI(Tl) crystals can be readily adapted for undoped CsI crystal growth.
Additionally, Bridgman growers, commonly used for CsI crystal growth, are available for this purpose.
In the KIMS experiment, the lowest background achieved at 2\,keV was about 2.5\,counts/kg/keV/day with the latest crystal delivered from Beijing Hamamatsu~\cite{Kim:2011af}.
The radioactive background in the PMTs and the direct attachment of PMTs to the crystal contribute to external background at low energy, estimated to be approximately 0.5--1\,counts/kg/keV/day~\cite{Lee:2007iq}.
Replacing the PMTs with SiPM arrays significantly reduces the external background contribution.
Indeed, the contributions from $^{134}$Cs, estimated at the level of 1.0\,count/kg/keV/day, can be considered negligible if crystals are grown using 20 years of underground-stored powder.
This is attributed to the relatively short half-life of $^{134}$Cs, which is only 2.1 years.
By leveraging existing powder and employing low-background crystal growing technology, the combination of undoped CsI crystals with SiPM arrays has the potential to achieve a background level of less than 1\,count/kg/keV/day.

We are using a conservative estimate of 26.2\,PE/keV for light yields with the SiPM arrays, even though there is an around 10\% increase in light output at 77\,K compared to 86\,K.
In the low-energy data acquisition setup, where two SiPM arrays are attached to two ends of the crystal, we can impose a requirement for coincident photoelectrons in the two SiPM arrays to minimize noise triggering while achieving the minimum thresholds, corresponding to an energy threshold of roughly 0.08\,keV.
In contrast to PMT measurements, where PMT-induced noise from Cherenkov radiation of charged particles passing through quartz or glass materials can dominate low-energy events and increase the analysis threshold~\cite{COSINE-100:2020wrv}, such noise is less frequent in SiPM arrays since quartz or glass materials can be minimized in their construction.
While DCR could be a potential concern, they are sufficiently low at 77\,K.
For the purpose of calculating dark matter detection sensitivities, we assume a 5\,PE threshold, consistent with CsI(Na) measurements using PMT readout~\cite{Akimov:2017ade}.

The proposed 200\,kg undoped CsI experiment can be implemented using a 5$\times$5 array of 8.7\,kg CsI modules, each with dimensions of 8\,cm $\times$ 8\,cm $\times$ 30\,cm (similar to the size used in the KIMS experiment~\cite{KIMS:2007wwj,sckim12}).
The two ends of the 8\,cm $\times$ 8\,cm side can be covered with 13$\times$13 SiPM arrays with 6\,mm $\times$ 6\,mm cells.
The assumed parameters for the analysis include a flat background of 1\,count/kg/keV/day, a light yield of 26.2\,PE/keV, corresponding to 0.08\,keV, and an analysis threshold of 5\,PE.
Additionally, an increased light yield of 30\,PE/keV, achieved through the application of a wavelength shifter~\cite{Wang:2022ekc}, corresponding to a 0.07\,keV threshold, is considered in the evaluation.
For the benchmark model of the low-mass dark matter search, low-energy signals via Migdal electrons~\cite{Migdal,Ibe:2017yqa} induced by WIMP-proton spin-dependent interaction are assessed.

To estimate the dark matter signal enhancement that is provided by the Migdal effect, we generate expected signals by multiplying the nuclear recoil rate of WIMP interactions~\cite{Savage:2008er} and ionization probability of the Migdal effect~\cite{Ibe:2017yqa,COSINE-100:2020wrv}. For various WIMP masses and WIMP-proton spin-dependent interactions, we simulate the expected event rates in the specific context of the standard halo model~\cite{Lewin:1995rx,RevModPhys.85.1561}.
Responses that include form factors and proton spin values of the nuclei are implemented from the publicly available  {\sc dmdd} package~\cite{dmdd,Gluscevic:2015sqa}. 
Due to the absence of calculations for cesium, we utilized the ionization probability from iodine. Additionally, we experimented with  xenon's ionization probability  to cross check the results.
The ionization probabilities from elements with similar atomic mass numbers to cesium yielded comparable results.
Assuming a one-year data exposure with 200\,kg of undoped CsI crystals, the Poisson fluctuation of the measured number of photoelectrons is taken into account for the detector resolution, as discussed in Ref.~\cite{Ko:2022pmu}.
The nuclear recoil quenching factors (QFs) which is the ratio of the scintillation light yield from cesium or iodine recoil relative to that from electron recoil for the same energy are additionally considered.
The measured quenching factor of undoped CsI from Ref.~\cite{Lewis:2021cjv} is incorporated into the calculations. In this calculation, detection efficiency and systematics are not considered.

We use an ensemble of simulated experiments to estimate the sensitivity of the experiment.
This ensemble test quantifies the expected cross-section limits for WIMP-proton spin-dependent interactions with the Migdal effect.
For each experiment, we determine a simulated spectrum based on a background-only hypothesis, with the assumed background achieved through Gaussian fluctuations.
The simulated data is then fitted using a signal and background hypothesis.
The fit is performed within the energy range of 5\,PE to 400\,PE for each WIMP model with various masses, employing the Bayesian approach used in the COSINE-100 experiment~\cite{COSINE-100:2021xqn}.
The 1$\sigma$ and 2$\sigma$ standard-deviation probability regions of the expected 90\% confidence level limits are calculated from 1000 simulated experiments.

\begin{figure}[htb]
    \begin{center}
    \includegraphics[width=1.0\columnwidth,keepaspectratio]{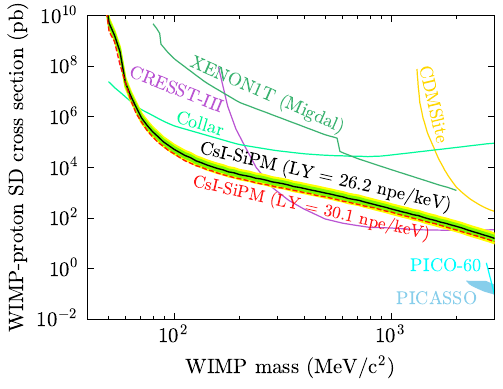}
    \end{center}
    \caption{The expected 90\% confidence level limits for the undoped CsI with a 200\,kg, 1-year exposure using the Migdal effect on the WIMP-proton spin-dependent cross-section are presented, assuming the background-only hypothesis. In addition to the nominal assumed light yield of 26.2\,PE/keV (black solid line with 1-$\sigma$ (green) and 2-$\sigma$ (yellow) systematic bands), higher light yields, assuming the application of the wavelength shifter of 30\,PE/keV (red dashed line), are evaluated. These limits are compared with the current best limits from XENON1T (Migdal)\cite{xenoncollaboration2019}, CRESST-\MakeUppercase{\romannumeral 3} LiAlO$_{2}$~\cite{CRESST:2022dtl}, Collar\cite{collar2018}, CDMSlite~\cite{supercdmscollaboration2018}, PICO-60~\cite{picocollaboration2019}, and PICASSO~\cite{behnke2017}.
    }
    \label{MDsensitivity}
\end{figure}

The limits on the WIMP-proton spin-dependent interaction with the Migdal effect shown in Fig.~\ref{MDsensitivity} are compared with the current best limits on the low-mass WIMP searches from XENON1T with Migdal\cite{xenoncollaboration2019}, CRESST-\MakeUppercase{\romannumeral 3} LiAlO$_{2}$~\cite{CRESST:2022dtl}, Collar~\cite{collar2018}, CDMSlite~\cite{supercdmscollaboration2018}, PICO-60~\cite{picocollaboration2019}, and PICASSO~\cite{behnke2017}.
Taking advantage of odd-proton numbers and low-energy threshold, the projected sensitivity with 200\,kg of undoped CsI experiment can reach the world's best sensitivities for WIMP masses between 60\,MeV/c$^{2}$ and 2\,GeV/c$^{2}$.

\section{Conclusion}
The characteristics of a pure CsI crystal coupled with two SiPMs were investigated at temperatures ranging from 86\,K to room temperature.
The increased light yield and decay time at low temperatures were observed consistently with the literature.
We obtained a light yield of 26.2 $\pm$ 1.3 PE/keV at 86\,K, which is 12 times larger than the light yield at room temperature.
Given our capacity for producing low-background undoped CsI crystals, we investigated the expected sensitivity of a 200\,kg undoped CsI experiment operating at 77\,K with a 1-year exposure, 1\,count/kg/keV/day background rate, and a 5\,PE energy threshold.
In this scenario, the undoped CsI crystal can explore low-mass dark matter between 60\,MeV/c$^{2}$ and 2\,GeV/c$^{2}$ with the world's best sensitivities for WIMP-proton spin-dependent interactions.

\section*{Acknowledgements}

This work was supported by the Institute for Basic Science (IBS) under project code IBS-R016-A1.




\bibliographystyle{elsarticle-num}

\end{document}